\documentstyle[preprint,aps]{revtex}
\def\journal#1, #2, 1#3#4#5#6, #7    {
    {\sl #1~}{\bf #2}, #7    (1#3#4#5)#6}
\def\pr{\journal Phys. Rev., }

\def\prl{\journal Phys. Rev. Lett., }

\def\np{\journal Nucl. Phys., }
\def\pl{\journal Phys. Lett., }

\def\mpl{\journal Mod. Phys. Lett., }

\def\jmp{\journal J. Math. Phys., }
\def\jp{\journal J. Phys., }
\newcommand{\beq}[1]{\begin{equation}\label{#1}}
\newcommand\eeq{\end{equation}}
\newcommand{\ba}[1]{\begin{eqnarray}\label{#1}}
\newcommand{\baa}{\begin{eqnarray}}
\newcommand\ea{\end{eqnarray}}
\newcommand{\bee}{\begin{equation}}
\def\nn{\nonumber \\}
\def\l{\lambda}

\def\f{\varphi}

\def\tr{\tilde{\rho}}
\newcommand{\cl}{Calogero}
\newcommand{\h}{Hamiltonian}
\newcommand{\col}{collective}

\newcommand{\B}[1]{{\bf #1}}
\begin{document}
\draft
\title{Calogero-Sutherland model from excitations of Chern-Simons vortices}
\author{Ivan Andri\'c, Velimir Bardek, and Larisa Jonke
\footnote{e-mail address: 
andric@thphys.irb.hr \\ \hspace*{3cm} bardek@thphys.irb.hr \\ \hspace*{3cm} 
larisa@thphys.irb.hr }}
\address{Theoretical Physics Division,\\
Rudjer Bo\v skovi\'c Institute, P.O. Box 1016,\\
10001 Zagreb, CROATIA}
\maketitle
\begin{abstract}
\widetext
We consider a large-N Chern-Simons theory for the attractive bosonic 
matter (Jackiw-Pi model) in the \h , \col-field 
approach based on the 1/N expansion. We show that the dynamics of density 
excitations around the ground-state semiclassical configuration is 
governed by the \cl \ or by the Sutherland \h , depending on the symmetry 
of the underlying static-soliton configuration. The relationship between 
the Chern-Simons coupling constant $\l$ and the 
\cl-Sutherland statistical parameter $\l_c$ 
signalizes some sort of statistical transmutation accompanying the 
dimensional reduction of the initial problem.
\narrowtext
\end{abstract}
\vspace{1cm}
\hspace{1.7cm} PACS number(s): 11.10.Lm 74.20.Kk 03.65.Sq 05.30.-d 

\newpage
Gauge models of a scalar field with the Chern-Simons term\cite{100} in $2+1$ 
space-time dimensions are known to support soliton or vortex solutions\cite
{111,222}. By using the nonrelativistic field theory of the self-attracted 
bosonic matter minimally coupled to an Abelian Chern-Simons gauge field, the 
authors of Ref.\cite{111} have shown that there exists a static self-dual 
soliton solution for a specific choice of the coupling constant. We have 
rederived this soliton solution in the \col-field approach by including 
higher-order terms in the $1/N$ expansion\cite{4}. In our approach, this 
soliton saturates the Bogomol'nyi bound and does not receive quantum 
corrections to its energy in the next-to-leading approximation.
Furthermore, owing to the fact that our soliton is normalized to a large 
number $N$, being the number of bosonic particles, we can describe it as a 
spiky distribution effectively representing one-dimensional bubbles living on 
the $xy$ plane. This observation substantially simplifies the problem of 
quantum excitations about such a configuration and allows us to identify their 
dynamics with that of the \cl-Sutherland model. 

There exist several recent papers that elucidate the connection between the 
Chern-Simons-based anyonic physics in the fractional quantum Hall effect (FQHE)
 and the \cl-Sutherland model.  

It was noted in\cite{18} that there was a close similarity between the \cl \
model\cite{1} and the system of anyons at the lowest Landau level in a strong 
external magnetic field. It was conjectured in Ref.\cite{9} and later proved
 in Ref.\cite{10} that the two systems were in fact equivalent.
The equivalence was demonstrated on the algebraic grounds, i.e., by finding
 a complex, Bargmann-Fock representation of the underlying operator algebras.

Furthermore, it was shown that the ground state of the \cl -Sutherland
(CS)\cite{2} model and the Laughlin state for the FQHE
coincided exactly in the narrow-cylinder geometry\cite{7,14}.
It was also shown that there was a similarity between the edge states of a 
non-narrow 
droplet of the FQHE and the chiral Tomonaga-Luttinger liquid whose
exponent was equal to that of the chiral-constrained CS model\cite{11,12}. 

Using the hydrodynamic \col-field theory, the authors of the Ref.\cite{8}
were able to show that the fermion correlation functions along the 
boundaries of the FQH droplet were interpolated to the correlation functions of 
the CS model as the droplet width was continuously narrowed.
Finally, the FQH effect and the CS model shared the same, infinite-dimensional
$W_{\infty}$ algebra\cite{15,16,17}.

In this paper we would like to extend this equivalence
to a completely different physical situation. Using the \col-field theory
approach, we show that the Jackiw-Pi model which describes nonrelativistic 
 anyons interacting via the $\delta $-function attractive potential 
undergoes a dinamical reduction in dimensionality, i.e., it reduces to a 
one-dimensional CS system.

The \h \ for $N$ spinless bosonic particles in the presence 
of the vortex of the strength $v$, 
located at the point $Z$\cite{4}, is 
\baa \label{hd}
H= && -2\sum_{i=1}^N\frac{\partial^2}{\partial z_i\partial\bar z_i}
+\frac{\l^2}{2}\sum_{i=1}^N\left|\sum_{j\neq i}^N\frac{1}{z_i-z_j}\right|^2
-\l\sum_{i\neq j}^N\frac{1}{z_i-z_j}\frac{\partial}{\partial \bar z_i} \nn 
&& +\l\sum_{i\neq j}^N\frac{1}{\bar z_i-\bar z_j}\frac{\partial}{\partial z_i} 
-2v\sum_{i=1}^N\frac{1}{\bar z_i-\bar Z}\frac{\partial}{\partial z_i}-v\l
\sum_{i\neq j}^N\frac{1}{z_i-z_j}\frac{1}{\bar z_i-\bar Z}+V  ,\ea
where the potential $V$ depends only on the position of particles.
 The complex numbers $z_i=x_i+iy_i$ represent the position of the i-th particle.
>From the bosonic wave function we have extracted the prefactor given by 
$\prod_{i=1}^N(\bar z_i-\bar Z)^v$, which represents the afore-mentioned 
vortex.
We have already shown in\cite{4} that, in the leading order in N, the 
collective motion of the system is given by the classical solution
of \col-field theory. For the sake of clarity, we review this part
of Ref.\cite{4} once again. 
The \col-field approach to the Jackiw-Pi 
 anyonic system is described by the Hamiltonian
\ba 1
H=&& \frac{1}{2}\int d^2\B r\rho(\B r)\left\{ \B\nabla\pi(\B r)+\hat{n}
\times\left[\frac{1}{2}\frac
{\B\nabla\rho(\B r)}{\rho(\B r)}+\l\int d^2\B r'\rho(\B r')
\frac{\B r-\B r'}{|\B r-\B r'|^2}-v\frac{\B r-\B R}{|\B r-\B R|^2}
\right]\right\}^2 \nn 
+ && \l\pi\int d^2\B r\rho^2(\B r)+V, \ea
where the dimensionless constant $\l$ is the so-called statistical parameter
which is to tune the desired statistics, and 
$\hat{n}$ is the unit vector perpendicular to the plane in which particles 
move. The \col \ field $\rho(\B r)$  is the 
continuum limit of the dynamical quantity:
\beq 2
\rho(\B r)=\sum_{i=1}^N\delta(\B r-\B r_i), \eeq
where $\B r_i$ are the positions of N bosonic particles interacting 
through the long-range statistical Bohm-Aharonov-like vector potential
\beq 3
\B A(\B r)=\l \hat{n}\times\int d^2\B r'\rho(\B r')\frac{\B r-\B r'}
{|\B r-\B r'|^2} . \eeq 
The operator $\pi(\B r)$ is the canonical conjugate of the field $\rho(\B r)$:
\beq 4
[\B\nabla\pi(\B r),\rho(\B r')]=-i\B\nabla\delta(\B r-\B r') .\eeq
The $v$-dependent term reflects the vortex-type singularity which should be 
canceled by the $\B\nabla\ln\rho $ term at the point of the vanishig density 
$\rho(\B r)$, i.e., at $\B r=\B R$.

If we had extracted the prefactor $\prod_{i=1}^N (z_i-Z)^v$ from the particle
wave function, we would have obtained the same \h \ , the only difference being 
in the sign of the $\l$-dependent terms. It will later become apparent that 
this form of the effective \h \ describes vortices with the negative 
statistical parameter $\l$.  

If we fine-tune the coupling $g$ of the $\delta$-function potential $V$
\bee V=-g\sum_{i,j}^N\delta(z_i-z_j) \eeq
and choose $ g=|\l|$,  then the \h \ (\ref{1}) becomes
\bee \label{1'}
H= \frac{1}{2}\int d^2\B r\rho(\B r)\left\{ \B\nabla\pi(\B r)+\hat{n}
\times\left[\frac{1}{2}\frac
{\B\nabla\rho(\B r)}{\rho(\B r)}+|\l|\int d^2\B r'\rho(\B r')
\frac{\B r-\B r'}{|\B r-\B r'|^2}-v\frac{\B r-\B R}{|\B r-\B R|^2}
\right]\right\}^2. \eeq
The leading part of the \col -field \h \ in the $1/N$ expansion is 
given by the effective potential
\bee \label{veff}
V_{\rm eff}=\frac{1}{2}\int d^2\B r\rho(\B r)\left[\frac{1}{2}\frac
{\B\nabla\rho(\B r)}{\rho(\B r)}+|\l|\int d^2\B r'\rho(\B r')
\frac{\B r-\B r'}{|\B r-\B r'|^2}-v\frac{\B r-\B R}{|\B r-\B R|^2}
\right]^2. \eeq
Owing to the positive definiteness of the effective potential (\ref{veff}), 
the Bogomol'nyi limit appears. The Bogomol'nyi bound is saturated by the 
 positive normalizable solution $\rho_0(\B r)$ of the equation
\beq 5
\frac{1}{2}\frac{\B\nabla\rho_0(\B r)}{\rho_0(\B r)}+|\l|\int d^2\B r'
\rho_0(\B r')\frac
{\B r-\B r'}{|\B r-\B r'|^2}-v\frac{\B r-\B R}{|\B r-\B R|^2}=0 .\eeq

Let us now in more detail examin the static solutions of the Bogomol'nyi 
equation (\ref{5}) and the corresponding excitations whose dynamics 
will be shown to be equal to that of the CS model.
In the rectangular coordinates $(x,y)$, Eq. (\ref{5}) can be written in the form
\begin{mathletters}\label{a}
\beq 6
\frac{1}{2}\frac{\partial}{\partial x}\ln\rho_0(\B r)+|\l|\int d^2\B r'\rho_0
(\B r')\frac{x-x'}{|\B r-\B r'|^2}-v\frac{x-X}{|\B r-\B R|^2}=0 ,\eeq 
\beq 7
\frac{1}{2}\frac{\partial}{\partial y}\ln\rho_0(\B r)+|\l|\int d^2\B r'\rho_0
(\B r')\frac{y-y'}{|\B r-\B r'|^2}-v\frac{y-Y}{|\B r-\B R|^2}= 0.\eeq
\end{mathletters}
There exists an interesting solution to this coupled set of equations, 
depending only on one variable, let us say $x$, for definiteness. 
Since the integral 
kernel in the second equation is an odd function in $y-y'$, the set (\ref{a}) 
is consistent only for $|\B R|\rightarrow\infty$.
The only relevant equation is therefore given by
\beq 8
\frac{1}{2}\frac{\partial}{\partial x}\ln\rho_0(x)+|\l|\pi\int dx'\rho_0(x')
{\rm sign}(x-x')=0 ,\eeq
where we have used the result valid for infinite space:
\bee  \int\frac{dy'}{(x-x')^2+(y-y')^2}=\frac{\pi}{|x-x'|} ,\;\; x\neq x' .\eeq
The integro-differential equation (\ref{8}) can be reduced to a differential 
one by taking the derivative with respect to $x$:
\beq 9
\frac{1}{2}\frac{\partial^2}{\partial x^2}\ln\rho_0(x)+2|\l|\pi\rho_0(x)=0. 
\eeq
This equation has a positive and normalizable solution given by
\beq r
\rho_0(x)=\frac{N\kappa}{\cosh^22\kappa x},\;\;\kappa=\frac{|\l|\pi N}{2} .\eeq

It is interesting to note that our soliton solution (\ref{r}) can be obtained 
as a special case of the general solution to the Liouville equation. 
In fact, applying the gradient operator, we can transform Eq. (\ref{5}) into
the Liouville equation:
\bee \label{lx} 
\frac{1}{2}\Delta\ln\rho_0(\B r)+2|\l|\pi\rho_0(\B r)=0 . \eeq
It is known that the general, positive solution to this equation is given by
\beq p
\rho_0(\B r)=\frac{2}{|\l|\pi}\frac{\left|\frac{df(z)}{dz}\right|^2}{
\left[ 1+|f(z)|^2\right]^2} , \eeq
where $f(z)$ is an arbitrary holomorphic function of $z=x+iy$, but chosen so 
that $\rho_0(\B r)$ is nonsingular and nonvanishing except 
at infinity. It is easy to see that the only choice for $f(z)$ which generates 
a single-variable, $x$-dependent solution is given by 
\bee
f(z)={\rm e}^{az} , \eeq
where $a$ is an arbitrary real constant. The requirement of normalizability
fixes the constant to be $a=2\kappa$. This finally reproduces our solution 
(\ref{r}). 

The \col-field configuration (\ref{r}) describes the ground state of N 
bosonic particles with
the attractive $\delta-$function interaction, as can be easily seen from the 
corresponding one-dimensional effective potential
\beq w
V_{\rm eff}=\frac{1}{2}\int dx \rho(x)\left[\frac{1}{2}\frac{\partial}
{\partial x}
\ln\rho(x)+|\l|\pi\int dx'\rho(x'){\rm sign}(x-x')\right]^2 .\eeq
Actually, the $\delta$ interaction appears as the cross term in the square 
(\ref{w}):
\ba q
&&V_{\rm eff}=\frac{1}{8}\int dx \frac{1}{\rho(x)}\left(\frac
{\partial\rho(x)}{\partial x}\right)^2-|\l|\pi\int dx\rho^2(x) 
\nn
&&+\frac{\l^2\pi^2}{2}\int 
dx \rho(x)\left[\int dx'\rho(x'){\rm sign}(x-x')\right]^2 .\ea
Using the identity
\bee {\rm sign}(x-y){\rm sign}(x-z)+{\rm sign}(y-x){\rm sign}(y-z)+
{\rm sign}(z-x){\rm sign}(z-y)=1,\eeq one can show that the contribution of 
the last term in (\ref{q}) transforms into an irrelevant constant which 
only shifts 
the zero point of the energy scale. With increasing number of particles N, 
the soliton profile of width proportional to $1/N$ (\ref{r}) 
becomes thinner, finally taking the form of 
the $\delta$ distribution:
\beq g \rho_0(x)=N\delta(x) . \eeq
This can be readily obtained by using one of the appropriate representations
 of the $\delta$ function:
\bee \delta(x)=\lim\limits_{\epsilon\to 0}\frac{{\rm exp}(x/\epsilon)}
{\epsilon[1+
{\rm exp}(x/\epsilon)]^2}\;,\;\epsilon=\frac{1}{4\kappa} .\eeq
  Consequently, particles  are restricted by their 
statistical interaction (effectively, the attractive $\delta$-function
interaction) to move along the $y$ axis. Although the motion of particles takes 
place in a two-dimensional space, the system is effectively one-dimensional.
The only relevant degree of freedom we are left with can be described by 
the residual \col-field excitation $\tr(y)$, which also lives on the 
$y$ axis, i.e.,
\beq e
\rho(\B r)=\delta(x)\tilde{\rho}(y). \eeq
Having written the excited \col-field configuration in the form (\ref{e}), 
we automatically normalize the residual field $\tr$ to the same number 
of particles $N$.

To find the dynamics of this excitation, we must insert the factorization
 form (\ref{e}) into the \col \ \h \ (\ref{1}).
A simple calculation yields
\ba h
&& H=\frac{1}{2}\int dx dy \rho_0(x)\tr(y)\left[\frac{\partial\pi}{\partial x}-
\frac{1}{2}\frac{\partial}{\partial y}\ln\tr(y)-|\l|\int dx'dy'\rho_0(x')
\tr(y')\frac{y-y'}{(x-x')^2+(y-y')^2}\right]^2 \nn
&& +\frac{1}{2}\int dx dy \rho_0(x)\tr(y)\left[\frac{\partial\pi}{\partial y}+
\frac{1}{2}\frac{\partial}{\partial x}\ln\rho_0(x)+|\l|\int dx'dy'\rho_0(x')
\tr(y')\frac{x-x'}{(x-x')^2+(y-y')^2}\right]^2 .\ea

If we take into account the soliton equation (\ref{9}) and the limiting form 
of the corresponding solution (\ref{g}), we can show that 
the \col \ \h \ is
\beq z
H=\frac{1}{2}\int dy\tr(y)\left(\frac{\partial\pi}{\partial y}\right)^2+
\frac{1}{2}\int dy\tr(y)\left[\frac{1}{2}\frac{\partial}{\partial y}\ln
\tr(y)+|\l|\int dy'\frac{\tr(y')}{y-y'}\right]^2 .\eeq
Here we have neglected the x-dependence of the conjugate momentum $\pi$ since
all particles are allowed to move only along the $y$ axis.
By rescaling the field $\tr(y)\to c\rho(y)$ and the momentum $\pi(y)\to 
\pi(y)/c$, we can recast the \col \ \h \ (\ref{z}) into the \cl \ 
form\cite{3,6} as
\beq c
H=\frac{1}{c}\left\{\frac{1}{2}\int dy\rho(y)\left(\frac{\partial\pi}{\partial 
y}\right)^2+\frac{1}{2}\int dy\rho(y)\left[\frac{\l_c-1}{2}\frac{\partial}
{\partial y}\ln\rho(y)+\l_c\int dy'\frac{\rho(y')}{y-y'}\right]^2 \right\} ,\eeq
where the constant c, the anyonic parameter $\l$ and the \cl \ statistical 
parameter $\l_c$ are interrelated by
\beq o
 c=\l_c-1\;{\rm and}\;|\l| c^2=\l_c , \eeq
finally leading to the relation \bee \l_c=|\l|(\l_c-1)^2 . \eeq
It is interesting to observe that for a fixed value of the anyonic 
statistical parameter $\l$ there are, in principle, two different 
values of the corresponding \cl-Sutherland statistical 
parameter $\l_c^+$ and $\l_c^-$ connected by the relation $\l_c^+\l_c^-=1$.
This relation somehow reflects the duality of the $\l_c$ and $1/\l_c$ 
\cl -Sutherland models.

Now we are going to show that our system of Jackiw-Pi anyons can be similarly 
reduced to the Sutherland model. In this case, we are looking for a 
radially symmetric soliton solution to Eq. (\ref{5}), describing the 
vortex located at the origin.
It has been shown in\cite{4} that there exists a radially symmetric, 
positive and normalizable \col -field configuration which minimizes the 
energy (\ref{1}). It is given by the vortex form
\beq y
\rho_0(r)=\frac{|\l| N^2}{2\pi r^2}\left[\left(\frac{r_0}{r}\right)^{
\frac{N|\l|}{2}}+\left(\frac{r}{r_0}\right)^{\frac{N|\l|}{2}}\right]^{-2} .
\eeq
The vorticity $v$ is fixed by the normalization condition and is given by
\bee v=N\frac{|\l|}{2}-1 .\eeq
The parameter $r_0$ reflects the scale invariance of the problem and cannot be
determined. Now, if N is large enough, we can again replace the soliton 
configuration $\rho_0(r)$ with the $\delta$ profile:
\beq n \rho_0(r)=\frac{N}{2\pi}\frac{\delta(r-r_0)}{r_0} .\eeq
The residual \col -field excitations $\tr(\f)$ can move only along the circle
 of radius $r_0$. The corresponding \col \ \h \ can be found along similar 
lines, 
explicitly given for the rectangular geometry:
\baa \label{u}
&&H=\frac{1}{2}\int rdr d\f\rho_0(r)\tr(\f)\left[\frac{\partial\pi}{\partial r}-
\frac{1}{2r}\frac{\partial}{\partial\f}\ln\tr(\f)-|\l|\int r'dr' d\f'\rho_0(r')
\tr(\f')\frac{r'\sin(\f-\f')}{|\B r-\B r'|^2}\right]^2 \nn 
&& +\frac{1}{2}\int rdr d\f\rho_0(r)\tr(\f)\left[\frac{1}{r}\frac
{\partial\pi}{\partial\f}+\frac{1}{2}\frac{\partial}{\partial r}\ln\rho_0(r)
-|\l|\int r'dr' d\f'\rho_0(r')
\tr(\f')\frac{r'\cos(\f-\f')-r}{|\B r-\B r'|^2}
-\frac{v}{r}\right]^2 .\ea
 Using the radial part of the Bogomol'nyi equation (\ref{5}),
\bee 
\frac{1}{2}\frac{\partial}{\partial r}\ln\rho_0(r)-
|\l|\int r'dr' d\f'\rho_0(r')\frac{r'\cos(\f-\f')-r}{|\B r-\B r'|^2}-
\frac{v}{r}=0, \eeq 
and the limiting form of the corresponding solution (\ref{n}),
it can be shown that the \col \ \h \ for two-dimensional anyons finally 
reduces to the \col \ Sutherland Hamiltonian\cite{19} 
\beq l
H=\frac{1}{2r_0^2}\int d\f\tr(\f)\left(\frac{\partial\pi}{\partial\f}\right)^2+
\frac{1}{2r_0^2}\int d\f\tr(\f)\left[\frac{1}{2}\frac{\partial}{\partial\f}
\ln\tr(\f)+\frac{|\l|}{2}\int d\f\tr(\f)\cot\frac{\f-\f'}{2}\right]^2 ,\eeq
up to the irrelevant constant term.
Further rescaling finally connects the parameter $\l$ with the \cl-Sutherland 
statistical parameter $\l_c$ by the same set of relations (\ref{o}).

It is evident from the relations (\ref{o}) that, for example,
 as anyonic statistics in two dimensions 
approaches "super" bosonic statistics ($|\l|\to 2$), 
 the corresponding statistics of the generated \cl \ model goes to
bosonic statistics ($\l_c\to 2$) or semionic one ($\l_c\to 1/2$).
Moreover, for $|\l|\to \infty$, the  statistical parameter of the \cl \ model
approaches unity and we recover the \col -field \h \ for the $c=1$
matrix model.
This is the signal of some sort of statistical transmutation accompanying 
the dimensional reduction of the anyonic system.
However, the critical value
$|\l|=0$ is forbidden because, in this case, there would be no
static-soliton solution to Eq. (\ref{9}), which represents the
cornerstone of dimensional reduction.

Further study is still needed to fully understand the physical meaning 
of this dimensional reduction  and the 
statistical transmutation associated with it.


{\bf Acknowledgment}

This work was supported by the Scientific Fund of the Republic of Croatia.

\end{document}